\documentclass[aps,prapplied,twocolumn,superscriptaddress,nofootinbib,a4paper,longbibliography]{revtex4-2}

\usepackage[pdftex]{graphicx}
\usepackage{amsmath}
\usepackage{verbatim}
\usepackage{xcolor}
\usepackage{bm}
\usepackage{txfonts}
\usepackage{siunitx}
\usepackage[colorlinks=true,allcolors=blue]{hyperref}

\begin{document}
\title{Beating ringdowns of near-degenerate mechanical resonances}
	
\author{Matthijs H.\ J.\ de Jong}
\affiliation{Department of Precision and Microsystems Engineering, Delft University of Technology, Mekelweg 2, 2628CD Delft, The Netherlands}
\affiliation{Kavli Institute of Nanoscience, Department of Quantum Nanoscience, Delft University of Technology, Lorentzweg 1, 2628CJ Delft, The Netherlands}

\author{Andrea Cupertino}
\affiliation{Department of Precision and Microsystems Engineering, Delft University of Technology, Mekelweg 2, 2628CD Delft, The Netherlands}

\author{Dongil Shin}
\affiliation{Department of Precision and Microsystems Engineering, Delft University of Technology, Mekelweg 2, 2628CD Delft, The Netherlands}
\affiliation{Department of Materials Science and Engineering, Delft University of Technology, Mekelweg 2, 2628 CD Delft, The Netherlands}

\author{Simon Gröblacher}
\affiliation{Kavli Institute of Nanoscience, Department of Quantum Nanoscience, Delft University of Technology, Lorentzweg 1, 2628CJ Delft, The Netherlands}

\author{Farbod Alijani}
\affiliation{Department of Precision and Microsystems Engineering, Delft University of Technology, Mekelweg 2, 2628CD Delft, The Netherlands}

\author{Peter G. Steeneken}
\affiliation{Department of Precision and Microsystems Engineering, Delft University of Technology, Mekelweg 2, 2628CD Delft, The Netherlands}
\affiliation{Kavli Institute of Nanoscience, Department of Quantum Nanoscience, Delft University of Technology, Lorentzweg 1, 2628CJ Delft, The Netherlands}

\author{Richard A.\ Norte}
\email{r.a.norte@tudelft.nl}
\affiliation{Department of Precision and Microsystems Engineering, Delft University of Technology, Mekelweg 2, 2628CD Delft, The Netherlands}
\affiliation{Kavli Institute of Nanoscience, Department of Quantum Nanoscience, Delft University of Technology, Lorentzweg 1, 2628CJ Delft, The Netherlands}
	
\begin{abstract}
Mechanical resonators that possess coupled modes with harmonic frequency relations have recently sparked interest due to their suitability for controllable energy transfer and non-Hermitian dynamics. Here, we show coupling between high Q-factor ($>\!\!10^4$) resonances with a nearly 1:1 frequency relation in spatially-symmetric microresonators. We develop and demonstrate a method to analyze their dynamical behavior based on the simultaneous and resonant detection of both spectral peaks, and validate this with experimental results. The frequency difference between the peaks modulates their ringdown, and creates a beat pattern in the linear decay. This method applies both to the externally driven and the Brownian motion (thermal) regime, and allows characterization of both linear and nonlinear parameters. The mechanism behind this method renders it broadly applicable to both optical and electrical readout, as well as to different mechanical systems. This will aid studies using near-degenerate mechanical modes, for e.g.~optomechanical energy transfer, synchronization and gyroscopic sensors. 
\end{abstract}

\date{\today}
\maketitle
\section{Introduction}
The dynamics of coupled resonators have been intensely studied from numerous perspectives over the last centuries. With the advent of ultra-high Q mechanical resonators (\cite{Norte2016,Tsaturyan2017,Shin2022,Bereyhi2022} as recent examples), the regime of linear coupling between resonances~\cite{Zanette2018,Rodriguez2016,Dolfo2018} has seen renewed interest, due to the sensitivity to small perturbations. High-Q resonators can be utilized for their long coherence- and lifetimes~\cite{Faust2013,Schneider2014,Wilson2015}, but they often feature spatial symmetries that naturally predisposes them to have (near-) degenerate or harmonically related eigenmodes (e.g.~\cite{Gloppe2014,Mercierdelepinay2018,Halg2022,Lagala2022}). In the case of trampoline membranes, the near degenerate eigenmodes make them ideal candidates for various schemes in optomechanics, such as synchronization and phonon lasing~\cite{Heinrich2011,Liao2019,Zhang2012,Sheng2020,Zhang2022}, or heat and energy transport~\cite{Xu2016,Fong2019,Sheng2021}. Other effects, such as mechanical squeezing~\cite{Woolley2014,Patil2015,Pontin2016,Nielsen2017} and noise cancellation~\cite{Caniard2007,Tsang2012,Mercierdelepinay2021,deJong2022a} have been shown in a variety of geometries. Mechanical systems with near-degenerate eigenmodes are an attractive platform for studying exceptional points, non-reciprocal coupling and other phenomena of non-Hermitian (open) systems~\cite{Wiersig2014,Lau2018,Mathew2020,delPino2022,Patil2022}. An exceptional point, for example, requires the frequencies of the eigenmode to be degenerate while the decay rates are opposite (e.g.~one mode is driven, the other decays). Additionally, near-degenerate mechanical resonances feature direct applications to sensors such as gyroscopes~\cite{Nitzan2015,Defoort2017}. 

Due to the development of suspended micro- and nano-scale resonators, some recent studies have focused on the nonlinear behavior of coupled mechanical modes. Coupled nonlinear modes~\cite{Antoni2012,Samanta2015,Cadeddu2016} can lead to stabilization~\cite{Antonio2012} and low-noise oscillators. Strong coupling~\cite{Eichler2012} can be used in mechanical signal processing, while coherent~\cite{Chen2017,Chandrashekar2021} and nonlinear decay paths~\cite{Polunin2016,Guttinger2017,Shoshani2017} could be leveraged for controlling energy transfer between modes~\cite{Yan2022}. 

In this work, we leverage our fabrication precision to design and study high Q-factor mechanical modes with a 1:1 frequency relation. Based on structures with spatial symmetries and modelling by finite element methods, we find pairs of modes whose shapes are identical except for a rotation by \SI{90}{\degree} in the plane of the suspended structures. Measurements of these modes show two peaks in the displacement spectrum that are separated in frequency by less than 6 parts per million (ppm). The energy decay shows two different decay rates, which we can extract without resolving the two spectral peaks. There is evidence that energy is exchanged between these two resonances, which indicates a coupling mechanism. We distinguish between the frequency splitting of the peaks due to fabrication imperfections and the frequency splitting due to this coupling using a characterization method based on the simultaneous, resonant detection of both peaks. The detector has a bandwidth larger than the total frequency difference, such that we measure the superposition of both peaks. Based on the frequency difference and relative amplitudes of the peaks, we observe a characteristic beating pattern by performing ringdown measurements. From these ringdowns, we can extract damping rates $\gamma_{1,2}$ of the seemingly uncoupled peaks (i.e.~decay follows the expected exponential trend). Due to the difference between the peak frequencies $\omega_{1,2}$, a characteristic beat pattern appears, with an amplitude inversely proportional to the relative amplitude difference between the two spectral peaks. Modelling of this effect provides evidence of a small coupling between the two (unhybridized) modes with rate $J$, of a frequency shift due to a small Duffing nonlinearity, and it allows us to investigate the resonator phase decoherence in the thermal (Brownian) motion regime.

\section{Results}
\begin{figure*}
\includegraphics[width = \textwidth]{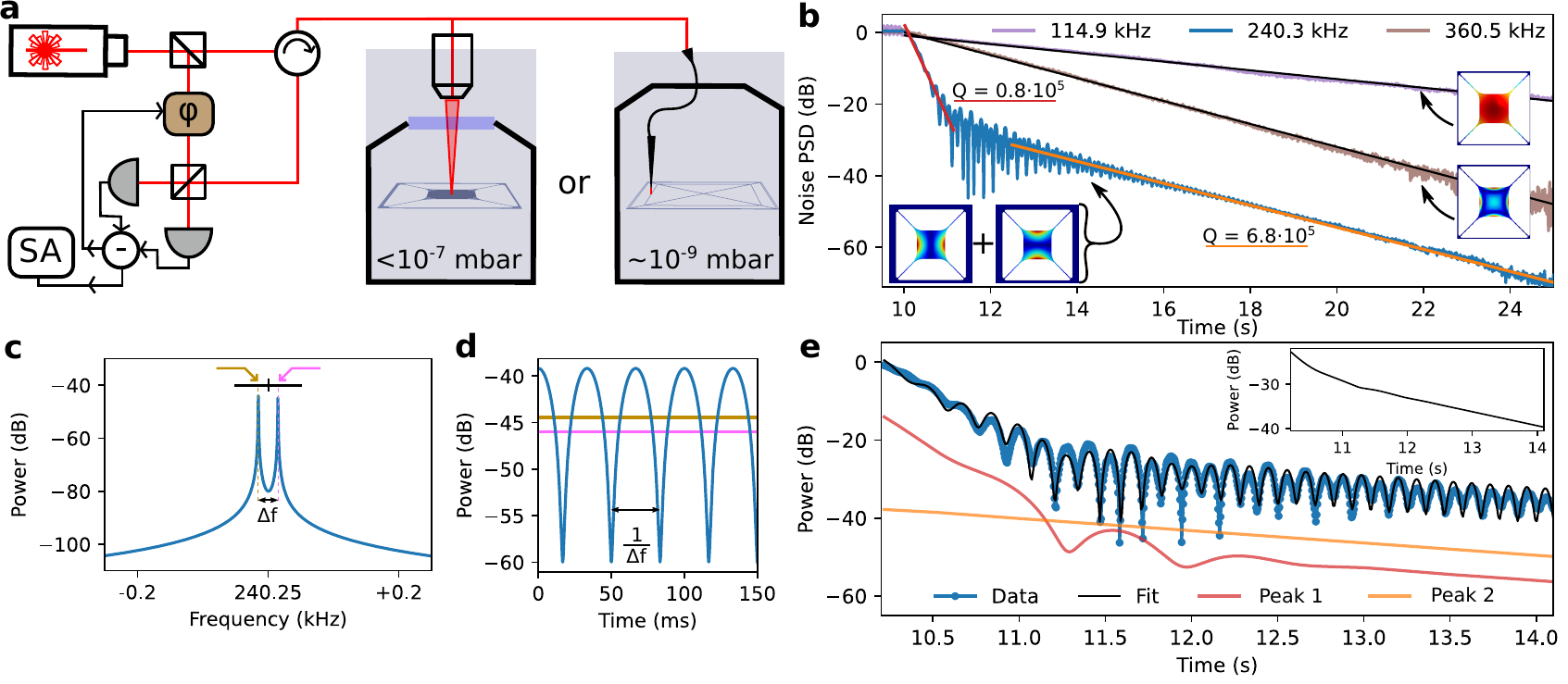}
\caption{\textbf{a}: Schematic of the two homodyne detection setups used to measure the mechanical motion of membranes (left) or spiderwebs (right). SA: Spectrum analyzer, $\phi$ phase shifter. The sketched laser positions match experimental conditions. \textbf{b}: Ringdowns of first three membrane modes (inset: mode shape, red indicates maximum displacement). They are aligned to the time where the driving is stopped, at $t = 10$~\si{\second}. The fundamental (purple) and third (brown) mode can be fit with single linear slopes (black), while the second mode shows two distinctly different linear slopes (red, orange, with fitted Q-factors) and a characteristic beating pattern. The spectrum of measured around this frequency shows two near-degenerate  peaks that we associate with the symmetric (simulated) mode shapes indicated in the inset. \textbf{c}: Simulated spectrum of two modes that exist as distinct spectral peaks separated by less than the detection bandwidth (black bar). \textbf{d:} The frequency difference $\Delta f$ leads to a characteristic beat in the ringdown signal. The relative amplitudes of the modes (gold, pink horizontal lines) determine the amplitude of the beating pattern. \textbf{e}: Simulated amplitudes of the near-degenerate peaks 1 and 2 based on extracted decay rates from \textbf{b}, reconstructed signal with the beating pattern (black) and the measured data (blue). Amplitudes are vertically offset for clarity. Inset shows the total power in the simulation, ignoring the detection efficiency. This is to verify that the energy monotonically decreases, as expected from a system with decay.}
\label{Fig1}
\end{figure*}

\paragraph*{Coupling and beating ---} \hspace{-1em}
We study the dynamics of near-degenerate mechanical modes using optical interferometers, shown schematically in Fig.~\ref{Fig1}\textbf{a}. Two types of resonators are studied, trampoline membranes~\cite{Norte2016} and spiderweb resonators~\cite{Shin2022}. A free-space readout system is used for trampoline membranes (Fig.~\ref{Fig1}\textbf{a}, left), while the narrow beams of the spiderwebs necessitate a lensed-fiber readout (Fig.~\ref{Fig1}\textbf{a}, right). Applying a resonant drive to a piezo shaker mounted on the sample holder excites mechanical modes of the resonator. By stopping the drive and letting the amplitude decay, we perform a ringdown experiment that allows precise measurement of the decay rate.

The amplitude of a linear harmonic oscillator decays exponentially, which we observe in the motion of the fundamental and third modes of the trampoline membranes (Fig.~\ref{Fig1}\textbf{b}, purple and brown lines). These fit well to a straight line (black) when plotted in log-scale. The ringdown trace at \SI{240}{\kilo\hertz} shows significantly different behavior, corresponding to two different slopes, two different decay rates. The extracted Q-factors (red, orange lines indicate fits) differ by almost an order of magnitude. We will show that this peculiar ringdown behavior is due to the near-degeneracy of two membrane modes (simulated shapes inset in Fig.~\ref{Fig1}\textbf{b}). This near-degeneracy is present for the second mode of the membranes due to symmetry, but absent for the fundamental and third modes. In the following, we refer to the two near-degenerate (second) modes as peaks with indices 1,2.

To model the ringdown with the near-degeneracy, we start with two (unhybridized) modes, $X_1 = \phi_1 (\vec{r}) x_1(t)$ and $X_2 = \phi_2 (\vec{r}) x_2(t)$, where mode shapes $\phi_1(\vec{r}),\phi_2(\vec{r})$ describe the spacial form of the mode through position vector $\vec{r}$, and generalized coordinates $x_1(t), x_2(t)$ describe the oscillating behavior in time. We obtain the mode shapes from finite-element simulations of our structure, and based on the symmetry of the structure and similarity of the two mode shapes, the effective masses of the unhybridized modes should be equal. We add to our model the decay rates $\gamma_1, \gamma_2$, resonance frequencies $\omega_1, \omega_2$ and a linear coupling with rate $J$~\cite{Zanette2018}. This coupling leads to hybridization, such that the eigenmodes of the system of equations are formed by linear superpositions of $x_1$ and $x_2$. This coupling $J$ could for instance occur via the substrate to which the resonator is anchored~\cite{Anetsberger2008,Truitt2013,deJong2022b}. In a ringdown measurement, we only observe the decay so our model does not need driving terms or noise sources; the initial amplitudes $x_1(t=0), x_2(t=0)$ of both unhybridized modes are non-zero. Thus we obtain the following set of equations of motion, 
\begin{equation}
\begin{aligned}
\ddot{x}_1 &+ \gamma_1 \dot{x}_1 + \omega_1^2 x_1 + J^2 x_2 = 0 \\
\ddot{x}_2 &+ \gamma_2 \dot{x}_2 + \omega_2^2 x_2 + J^2 x_1 = 0.
\label{EOM}
\end{aligned}
\end{equation}
It bears mention that the system of Eq.~\eqref{EOM} can always be diagonalized to obtain the eigenmodes with coordinates $y_1,y_2$ and frequencies $\omega_1', \omega_2'$, which are decoupled (see also the supplementary information (SI)~\cite{RRSI}\nocite{Virtanen2020,Landau1976,Hauer2013}, Sec.~A). Both descriptions yield the same dynamics if their parameters are properly matched. We will not diagonalize Eq.~\eqref{EOM} but keep coordinates $x_1, x_2$. This allows us to distinguish between the frequency difference $\omega_2 - \omega_1$ that comes from fabrication imperfections and the frequency difference due to coupling $J$. This will be required to separate the two contributions to the beating effect of Eq.~\eqref{Beating}, one from energy transfer and the other from the frequency difference. Furthermore, using the amplitudes of the unhybridized modes $x_1, x_2$ allows us to add a difference in detection efficiency to $x_1, x_2$ based on the mode shapes and position where the displacement is measured. Finally, we work in the weakly-coupled regime, so $x_1, x_2$ are not too different from $y_1, y_2$. 

The spectrum of our resonator around \SI{240}{\kilo\hertz} shows two distinct peaks separated by \SI{9}{\hertz}, as schematically shown in Fig.~\ref{Fig1}\textbf{c} (see also \cite{RRSI}, Sec.~B). They can be distinguished and fitted with a sum of two Lorentzians without overlap of the confidence interval of center frequency, so we refer to them as 'near-degenerate'. However, this frequency difference is smaller than the detection bandwidth of our spectrum analyzer during the ringdown measurement (SA in Fig.~\ref{Fig1}\textbf{a}), such that both peaks are captured in the single ringdown trace. This condition explains the observation of two different slopes in the same ringdown, as the two modes we observe have different decay rates. The slopes are each fitted to a single exponential decay. The mode with the fastest decay rate is (typically) more susceptible to the drive, such that its amplitude dominates the signal at the start of the ringdown measurement. For short timescales, we thus observe mainly the fast-decaying mode. At longer timescales, the slow-decaying mode dominates the signal, as the fast-decaying mode has lost most of its amplitude. This means that for long timescales, we observe mainly the slow-decaying mode. This leads to a single ringdown measurement showing two different slopes, and allows us to extract the decay rates of the two modes without spectrally resolving them.

The difference in Q-factor between the near-degenerate modes appears similar to the spread in Q-factors of fundamental modes of nominally identical devices (e.g.~\cite{Hoj2021,deJong2022b}). This spread is commonly attributed to local fabrication imperfections or material variations. Optical microscope images cannot always corroborate these explanations, since local material impurities (C, O, etc.) are too small to resolve optically. Both fabrication imperfections and material variations are local, and the two near-degenerate modes exist in the same device. However, the deflection profile is different, so the modes can be affected differently. This way, these effects could cause the difference in Q-factor between near-degenerate modes.

\begin{figure}
\includegraphics[width = 0.5\textwidth]{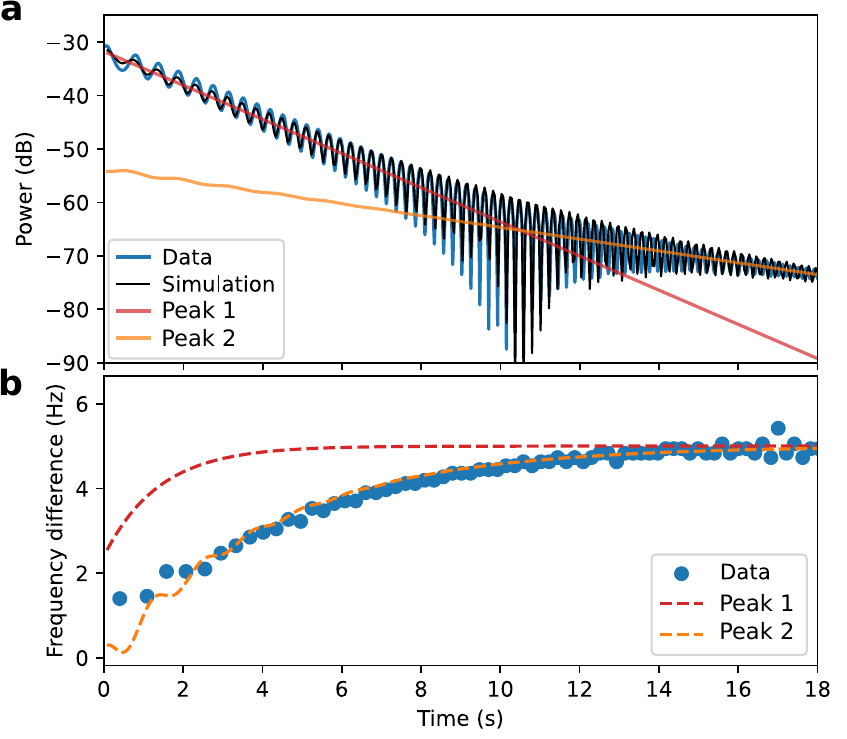}
\caption{\textbf{a}: Ringdown with a prominent beating pattern, measured (blue) and simulated (red, orange, black). This experiment is performed on a different device than those of Fig.~\ref{Fig1}, but in the same experimental setup. Two separate, linear slopes can be distinguished, but they are similar since the decay rates of both peaks are similar. \textbf{b}: Frequency difference between the two near-degenerate peaks as the ringdown progresses, determined from the distance between subsequent minima of \textbf{a}. The fit (orange) is an exponential described in the main text.}
\label{figfrequencyshift}
\end{figure}

In the kink between the two slopes (red, orange), we observe a particular beating pattern. Such patterns commonly indicate energy exchange~\cite{Zanette2018,Rodriguez2016}. However, the coupling rate necessary to create a beating pattern with the frequency we observe would mean that we are in the strong coupling regime. This regime is incompatible with the observation of the two (different) slopes, since strong coupling allows the energy to decay via the fastest-decaying mode, i.e.~one would only see the steepest slope (see \cite{RRSI} Sec.~C). Furthermore, the observed beating pattern is particular because it starts small in amplitude (at $t = 0$~\si{\second} in Fig.~\ref{figfrequencyshift}\textbf{a}), grows until some point ($t = 10.5$~\si{\second}) but then decreases in amplitude again ($t = 18$~\si{\second}). Beating patterns in ringdowns due to energy exchange typically show only a decrease in amplitude of the beat pattern~\cite{Guttinger2017,Shoshani2017}.

Instead, we propose a different effect that contributes to the beating pattern. The signal measured in a ringdown measurement is the total displacement of the resonator, which is the sum of the displacement of both coordinates $x_1$ and $x_2$ at the detection spot (Fig.~\ref{Fig1}\textbf{a}). The total detected displacement power is thus of the form
\begin{equation}
x^2 = \lvert e^{i\omega_1 t} \bar{x}_1 + e^{i\omega_2 t} d \bar{x}_2\rvert^2,
\label{Beating}
\end{equation} 
where we have split the coordinates $x_1, x_2$ into envelopes $\bar{x}_1, \bar{x}_2$, and fast-oscillating terms $e^{i\omega_1 t}, e^{i\omega_2 t}$. The bandwidth of our detector (\SI{100}{\hertz}) makes it much slower than the frequencies $\omega_1, \omega_2$ ($\simeq 240$~\si{\kilo\hertz}). We have assumed equal effective masses in Eq.~\eqref{EOM}, but in principle the effective masses are dependent on the position where the displacement is measured and the shape of the mode~\cite{Hauer2013}. To ensure energy conservation when matching our model with experimental observations, we must introduce a relative detection efficiency $d$. This constant factor is a function of the position where we measure the resonator displacement, and the two simulated mode shapes.

The frequency difference $\Delta_f = \omega_2 -\omega_1$ (gold, pink in Fig.~\ref{Fig1}\textbf{c,d}) creates a beat with period $1/\Delta f$, which is slow enough to be detected. This means the detected displacement power $x^2$ gains a periodic modulation, a beating which is proportional only to the frequency difference between the peaks and which does not imply energy transfer between them. This effect is well-understood as a beat between signals (see e.g.~Ref.~\cite{Cadeddu2016}), and it applies to different detection mechanisms (optical, electrical) since it only requires the bandwidth of the resonant detector to encompass both peaks. It is analogous to electronically or optically down-mixing a signal by sending in a different tone and observing the beat pattern. In this instance, it is passively obtained based on the small frequency difference between the peaks and the high Q-factor of the involved resonances.

To correctly fit the measured ringdown (Fig.~\ref{Fig1}\textbf{e}), we require both the linear coupling and the beating effect. Without the latter, the frequency of the beating pattern would indicate strong coupling, but this is incompatible with the presence of two slopes~\cite{Zanette2018}. Without the former, the beating pattern would only briefly appear in the ringdown as the peak amplitudes cross (\cite{RRSI} Sec.~C), whereas it extends much further in Fig.~\ref{Fig1}\textbf{e}. A nonlinear coupling ($\propto x^3$) would lead to a beating pattern that is qualitatively different to the one observed: the beating would slow down in frequency as the amplitude decays (see \cite{RRSI} Sec.~C for a comparison of the linear and nonlinear models). 

We obtain the individual center frequencies $\omega_1 = 2\pi\times 240,331.6 \pm 0.1$~\si{\hertz} and $\omega_2 = 2\pi\times 240,341.4 \pm 0.1$~\si{\hertz} of the peaks from a separate spectrum measurement (\cite{RRSI}, Sec.~B) which also allows us to estimate the detection efficiencies ($d \simeq 20$ in Fig.~\ref{Fig1}\textbf{e}). From the linear parts of the ringdown, we extract decay rates $\gamma_1 = 2\pi\times 6.0\pm 0.1$~\si{\hertz} and $\gamma_2 = 2\pi\times 0.707 \pm 0.001$~\si{\hertz}. We can simulate the ringdown using only three fit parameters: the initial positions $x_{10},~x_{20}$ and coupling rate $J$. The optimized fit (black) to the data (blue) is shown in Fig.~\ref{Fig1}\textbf{e}, with  dimensionless initial positions $x_{10} = 0.089\pm0.001$, $x_{20} = 0.071\pm 0.001$ and coupling $J/(2\pi) = 320\pm 5$~\si{\hertz}. The frequency shift due to the linear coupling, $J^2/\omega_1$, would correspond to \SI{0.4}{\hertz} if $\omega_1 = \omega_2$, so we are in the weakly-coupled regime. Nonetheless, there is sufficient coupling to observe energy exchange between the reconstructed powers of the unhybridized modes (red, orange lines in Fig.~\ref{Fig1}\textbf{e}). For example, between \SI{11.3}{\second} and \SI{11.5}{\second}, $x_1$ gains energy from $x_2$, shown by the increase of the reconstructed power. The power in $x_2$ correspondingly decreases, but this is not clearly visible in Fig.~\ref{Fig1}\textbf{e} due to the difference in detection efficiency. We numerically verify that the reconstructed power always decreases over time, as shown by plotting the total power in the inset of Fig.~\ref{Fig1}\textbf{e}. Our measurement thus marks the observation of purely linear coupling between two resonances with a near-degenerate frequency relation, without showing nonlinear power decay.

\paragraph*{Frequency shift ---} \hspace{-1em}
The beating pattern is periodic with the inverse of the frequency difference between the two near-degenerate peaks. Careful observation of Fig.~\ref{Fig1}\textbf{e} shows that this period is not constant; it changes with time. This means that the frequency difference is not constant, and that either one or both peaks change in frequency over time, similar to Ref.~\cite{Cadeddu2016}. In the previous section, the energy decay was observed to be linear, sufficiently so that no nonlinearity is necessary to describe the ringdown completely. However, a small nonlinearity might still be present and lead to a frequency shift without measurably affecting the energy decay. Since the period of the beating is inversely proportional to the frequency difference, analyzing this periodicity provides a measure sensitive to small nonlinearities.

From the ringdown shown in Fig.~\ref{figfrequencyshift}\textbf{a}, we extract the frequency difference by finding the minima of the beating pattern and taking the inverse of their time differences. We plot the frequency difference in Fig.~\ref{figfrequencyshift}\textbf{b} (blue), which shows an exponential trend towards $\Delta f = 5.0$~\si{\hertz}. It is likely that the drive pulled the frequencies of the two peaks together~\cite{Lifshitz2008}, which indicates the presence of a nonlinearity. Assuming a small Duffing nonlinear term ($\propto x^3$), the frequency shift due to frequency pulling is of the form~\cite{Lifshitz2008}
\begin{equation}
\omega_{nl} = \omega_0 + \frac{3}{8}\frac{\alpha}{m_\mathrm{eff}\omega_0} \tilde{x}^2,
\label{Duffingshift}
\end{equation}
with Duffing coefficient $\alpha$, effective mass $m_\mathrm{eff}$ and displacement amplitude $\tilde{x}$. 

\begin{figure*}[t]
\includegraphics[width = \textwidth]{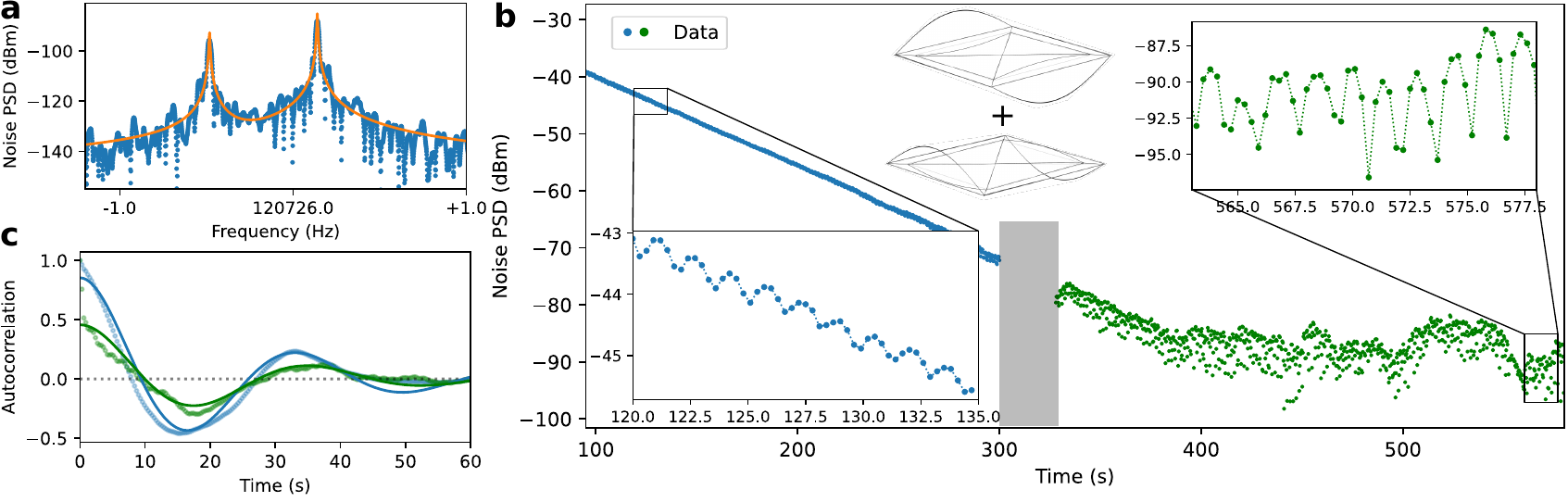}
\caption{\textbf{a}: Mechanical spectrum of near-degenerate modes of a spiderweb resonator (blue), with Lorentzian fits (orange). \textbf{b}: Two consecutive ringdowns performed on the same device, one showing the decay from the driven state (blue), the other showing the last part of the ringdown and subsequent transition towards a Brownian motion driven state (green). Insets highlight the beating behavior (dotted line drawn as a guide to the eye), which continues even in the Brownian-motion regime. Grey-colored region denotes the cut between the two ringdowns. The two simulated mode shapes are shown schematically. \textbf{c} Autocorrelation functions of the signal in the ringdown (blue) and Brownian-motion regime (green), showing a coherence time of $24 \pm 2$~\si{\second}. Points indicate the experimental data, solid lines indicate the fit.}
\label{figphasecoherence}
\end{figure*}

By observing the change in the frequency difference, we can extract which of the two peaks behaves nonlinearly, and whether it is softening or hardening. The linear parameters are extracted by the same procedure as before (the frequencies follow from a spectrum measurement, the decay rates are fits to the linear parts of the ringdown and their coupling is modelled via Eq.~\ref{EOM}). We find $\omega_1 = 2\pi \times 240,302.4\pm 0.2$~\si{\hertz}, $\gamma_1 = 2\pi \times 0.71 \pm 0.01$~\si{\hertz}, $\omega_2 = 2\pi \times 240,307.4\pm 0.1$~\si{\hertz}, $\gamma_2 = 2\pi \times 0.218 \pm 0.001$~ \si{\hertz}, and coupling $J = 2\pi \times 20 \pm 1$~\si{\hertz} (independent of the value of $\alpha$) with relative detection efficiency $d \simeq 50$. We calculate the frequency shift as the peak amplitude decays from Eq.~\eqref{Duffingshift}. The decay rates are different, which means that their frequency shift should change over time at different rates, illustrated by the dashed lines in Fig.~\ref{figfrequencyshift}\textbf{b}. The measured frequency shift is consistent with peak 2 being frequency-pulled by the drive. This implies a softening nonlinearity ($\alpha < 0$), which is usually associated with an external (e.g.~electrical or optical) force source (geometric nonlinearity tends to yield hardening behavior~\cite{Cadeddu2016,Samanta2018}). However, curvature of the membrane in the out-of-plane direction could also lead to a softening nonlinearity. Although the suspended membrane is nominally flat due to the tensile stress, some curving at the membrane edges can be observed under microscope (cf.~the supplementary of Ref.~\cite{deJong2023a}). In the absence of any external sources of (softening) nonlinearity, this curvature could be responsible for the observed change in periodicity of the beating pattern. Buckling\cite{Kanj2022} of the whole membrane is unlikely given the high (\SI{1}{\giga\pascal}) fabrication pre-stress. An estimate for $\alpha$ based on the onset of the Brownian motion regime yields a value $\alpha_2 = -3\times 10^{-21}$~\si{\per\square\meter\per\square\second}, but a more accurate value should be obtained after calibration of the displacement and detection efficiency. The nonlinear term is sufficiently small compared to the coupling $J$ that it is negligible for the measured ringdowns. Despite this, we can observe and measure the frequency shift due to the nonlinearity. Comparison to literature shows that we are three orders of magnitude more sensitive than Ref.~\cite{Cadeddu2016}, who utilize a similar technique but for GaAs nanowires. 

\paragraph*{Resonator decoherence ---} \hspace{-1em}
Frequency shifts of resonators due to external factors are associated with resonator phase decoherence~\cite{Schneider2014} or dephasing. This is detrimental for coherent control~\cite{Faust2013,Wilson2015}, which is important for many quantum mechanical applications. Classically, one can observe the spectral width integrated over sufficient time to obtain the phase decoherence time~\cite{Schneider2014}. We will describe how the beating effect provides another mechanism to evaluate the phase decoherence time by utilizing the second peak as a frequency reference. 

In spiderweb resonators~\cite{Shin2022}, there is a set of modes extremely close to degeneracy, which we observe as two peaks at frequencies $\omega_1 = 2\pi\times 120,725.517 \pm 0.002$~\si{\hertz}, $\omega_2 = 2\pi \times 120,726.141 \pm 0.001$~\si{\hertz} ($<6$~parts per million difference), as shown in Figs.~\ref{Fig1}\textbf{a} and \ref{figphasecoherence}\textbf{a}. These peaks have linewidths below the resolution of the spectrum analyzer in this setup (\SI{0.01}{\hertz}). We estimate their spectral linewidths to be \SI{2.5}{\milli\hertz} and \SI{2.3}{\milli\hertz} respectively (full width at half maximum, corresponding to $Q \simeq 25\times 10^6$), which matches to the Q-factor extracted from a linear fit to the ringdown measurement, $Q = 21.1\times 10^6$.

The ringdown measurement (Fig.~\ref{figphasecoherence}\textbf{b}), shows the beating pattern both in the strongly driven regime (blue) and in the Brownian motion regime (green). The dynamics of $x_1, x_2$ are much faster than our detection bandwidth, and we work at room temperature, so we cannot directly resolve the thermal decoherence rate~\cite{Wilson2015}. However, the beating pattern is sensitive to frequency fluctuations. A phase-shift in the beating pattern corresponds to a frequency shift between the two peaks. Thus monitoring the beating pattern phase allows us to observe frequency shifts, which correspond to phase decoherence between the two observed modes. We can do so by calculating the autocorrelation of the beating signal of Fig.~\ref{figphasecoherence}\textbf{b}, which we plot in in Fig.~\ref{figphasecoherence}\textbf{c}. It shows behavior similar to that of a single, underdamped particle on a spring undergoing Brownian motion, which has an autocorrelation given by~\cite{Reichl2016}
\begin{equation}
C(t) = \frac{\Gamma k_\mathrm{B} T}{m_\mathrm{eff}} e^{-\Gamma t} \left( \frac{1}{\Gamma} \cos(\delta t) - \frac{1}{\delta} \sin(\delta t)\right),
\end{equation}
with $\delta = \sqrt{\Omega^2 - \Gamma^2}$. Here the frequency $\Omega$ is the frequency of the beating pattern, and decay $\Gamma$ is related to the phase decoherence time, i.e.~when the amplitude of the autocorrelation drops by half. We fit a $1/\Gamma = 24.5\pm1.8$~\si{\second} ($24.8\pm2.8$~\si{\second}) in the driven (thermal) regime respectively. This is close to the linear (energy) decay time ($Q/\omega_{1,2} \simeq 27.8$~\si{\second}) extracted from the ringdown directly. This is the expected behavior for a linear harmonic oscillator: the phase decoherence time (measured via the beating pattern) should be similar to the energy decay time (measured directly from the ringdown). Additionally, we fit $\Omega = 0.195\pm0.001$~\si{\hertz} ($\Omega = 0.185\pm0.002$~\si{\hertz}) in the driven (thermal) regime, which matches well with the frequency difference between the two peaks $|\omega_1 - \omega_2|/\pi$. This illustrates that the beating effect can provide a measurement of the phase decoherence of (near-) degenerate modes of a mechanical resonator. 

\section{Conclusion}
We have experimentally studied high Q-factor mechanical modes of spatially-symmetric microresonators that are nearly degenerate. We provide evidence of a linear coupling between the two resonances, present without driving the resonator to nonlinearity. This is in contrast to previous studies, that generally require a quadratic or cubic term for coupling. To detect this linear coupling, we developed a characterization method based on a single, resonant detector. This detector has a detection bandwidth that encompasses both spectral peaks, such that we can see two decay rates in a single ringdown measurement. When the spectral peaks are sufficiently close in frequency, an interference effect occurs that leads to a particular beating pattern. The relative amplitudes of the individual peaks control the beating pattern amplitude, while their frequency difference controls the beating period. From the beating pattern, we can monitor the amplitudes of the individual spectral peaks and find their coupling rate. The beating period provides a sensitive measure to the relative frequency difference between the two peaks. This allows us to observe slight frequency shifts due to nonlinearities, which would be difficult to observe using other methods. Our Si$_3$N$_4$ trampoline membranes feature a softening nonlinearity, which we attribute to out-of-plane curving. The beating period can additionally be used to monitor frequency shifts associated with phase decoherence of the modes. In Si$_3$N$_4$ spiderweb resonators, we find the phase decoherence time is similar to the energy decay time, as expected for linear harmonic oscillators.

Our characterization method is applicable to both optical and electrical readout schemes, and is particularly suited to high Q-factor mechanical resonators. This type of resonators is highly relevant for e.g.~sensing, optomechanics and forms an attractive platform to study non-Hermitian systems. We have thus developed a broadly applicable method to characterize near-degenerate resonators, which can be immediately applied to studies in different fields. 

\vspace{0.25cm}
\paragraph*{Data availability}\mbox{}\\
All data, simulations, measurement and analysis scripts in this work are available at \href{https://doi.org/10.4121/21428517.v1}{https://doi.org/10.4121/21428517.v1}.

\vspace{0.25cm}
\paragraph*{Acknowledgments}\mbox{}\\
R.N. would like to acknowledge support from the Limitless Space Institute's I2 Grant. This publication is part of the project, Probing the physics of exotic superconductors with microchip Casimir experiments (740.018.020) of the research programme NWO Start-up which was partly financed by the Dutch Research Council (NWO).  M.J., A.C., and R.N. acknowledge valuable support from the Kavli Nanolab Delft and from the Technical Support Staff at PME, 3mE Delft.
\clearpage

\setcounter{figure}{0}
\renewcommand{\thefigure}{S\arabic{figure}}
\setcounter{equation}{0}
\renewcommand{\theequation}{S\arabic{equation}}
\section{Supplementary Information}
\subsection{Decoupling the EOM}
It is convenient to obtain the relation between the parameters of the coupled and decoupled equations of motion for our system of near-degenerate resonances. We start from the Fourier-transform of Eq.~\eqref{EOM}, where we have simplified $\omega_1 = \omega_0$, $\omega_2 = \omega_0 + \nu$ and $(\omega_0 + \nu)^2 \simeq \omega_0^2 + 2\nu\omega_0$ (thus assuming small $\nu$). Thus we start the equation of motion 
\begin{equation}
\begin{bmatrix}
\omega_0^2 + i\gamma_1 - \omega^2 &-J^2 \\
-J^2 &\omega_0^2 + 2\nu\omega_0 + i\gamma_2 - \omega^2
\end{bmatrix}
\begin{bmatrix}
x_1 \\
x_2
\end{bmatrix}
=
\begin{bmatrix}
0 \\
0
\end{bmatrix},
\end{equation}
which is of the form $\bm{\mathrm{A}}\vec{x} = \vec{0}$. It is straightforward to obtain the eigenvectors $\vec{e}_1,~\vec{e}_2$ of matrix $\bm{\mathrm{A}}$, and we put them in a $2\times 2$ matrix $\bm{\mathrm{U}} = [\vec{e}_1~\vec{e}_2]$. Then we can obtain the diagonalized matrix $\bm{\mathrm{B}}$ using $\bm{\mathrm{B}} = \bm{\mathrm{U}}^{-1}\bm{\mathrm{AU}}$. The coordinates in which the equations are decoupled are then given by $\vec{y} = \bm{\mathrm{U}}^{-1}\vec{x}$. The eigenvalues (diagonal elements) of $\bm{\mathrm{B}}$ are then given by 
\begin{equation}
\begin{aligned}
(\omega_{1,2}')^2 &= \omega_0^2 + \nu\omega_0 + \frac{i}{2}(\gamma_1 + \gamma_2) \pm \frac{\tau}{2}, \\
\tau &= \sqrt{4J^4 + 4 \nu^2\omega_1^2 - \gamma_1^2 -\gamma_2^2 + 2\gamma_1\gamma_2 + 4i\nu\omega_0(\gamma_1 + \gamma_2)}. 
\end{aligned}
\end{equation}
These $\omega_{1,2}'$ are the frequencies of the eigenmodes of the system, which are the frequencies we observe in the spectrum. If we measure the splitting $2\pi\Delta f = \Delta \omega = |\omega_1' - \omega_2'|$ and want to know the detuning of the original coordinates $x_1$ and $x_2$, $\nu = |\omega_1 - \omega_2|$ for a given coupling $J$, we use 
\begin{equation}
\nu = \frac{1}{2}\left( \Delta \omega^2 + i \gamma_1 - i\gamma_2\right) - \sqrt{\Delta \omega^2 (\omega_0^2 + i\gamma_1) - J^4}.
\end{equation}
which allows us to find the frequencies of the coupled equations, $\omega_{1,2}$, from the decoupled ones and coupling rate $J$.

\subsection{Spectrum and detection efficiency}\label{Spectrum}

\begin{figure}
\includegraphics[width = 0.5\textwidth]{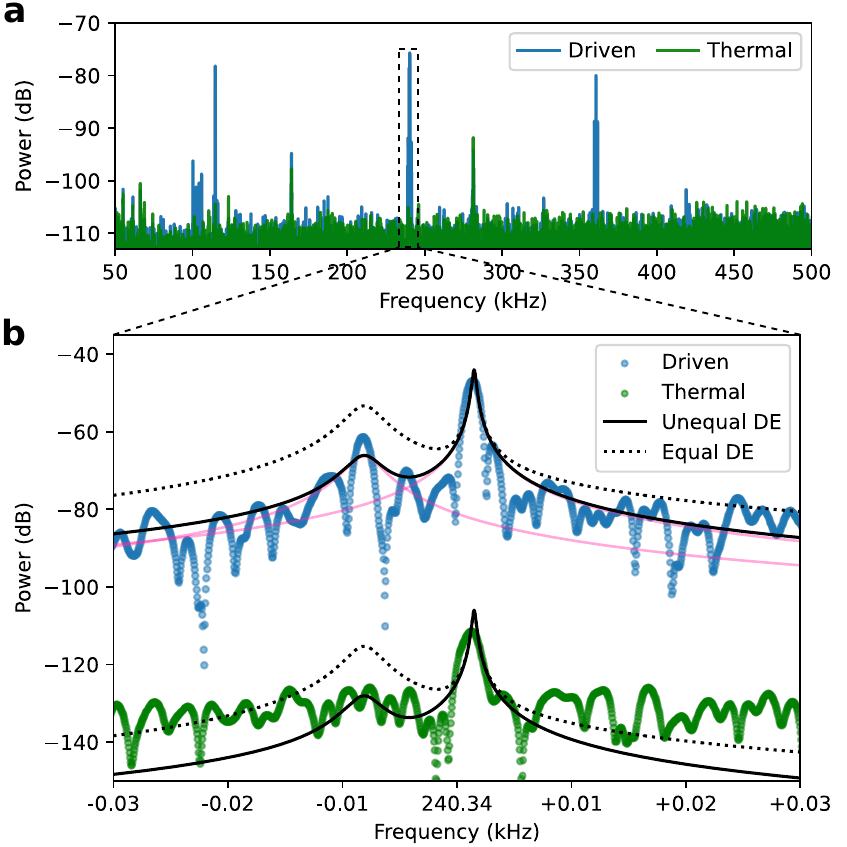}
\caption{\textbf{a}: Spectrum of a typical mechanical resonator, showing the first three modes (\SI{114.9}{\kilo\hertz}, \SI{240.3}{\kilo\hertz} and \SI{360.5}{\kilo\hertz} identified by their growth under white noise driving (blue) versus thermal driving (green). \textbf{b}: Spectrum around the near-degenerate peaks (vertically offset for clarity). Rounded peaks are the result of the \SI{1}{\hertz} measurement bandwidth used. Black line indicates Lorentzian fit of the sum of the peaks, dotted for equal detection efficiency and solid for the factor 20 less efficient detection of the lower-frequency mode. Pink lines indicate the individual Lorentzian components.}
\label{Figspectrum}
\end{figure}

A typical membrane device spectrum is shown in Fig.~\ref{Figspectrum}\textbf{a}. The the region of interest is around \SI{240.3}{\kilo\hertz}, where two very closely spaced peaks appear, separated by \SI{9}{\hertz}, as in Fig.~\ref{Figspectrum}\textbf{b}. Their frequency separation is well-resolved within the \SI{1}{\hertz} bandwidth of our spectrum analyser. The center frequencies of the peaks reported in the main text are extracted by fitting two summed Lorentzians to the spectrum, using the \texttt{curve\_fit} function from Scipy~\cite{Virtanen2020}. The laser spot of our interferometer is large enough to capture the majority of the membrane pad and thus detect both the near-degenerate mode shapes that we obtain from finite element method simulations. In the high-Q limit, the expected PSD for a harmonic oscillator is a Lorentzian around its resonance~\cite{Landau1976}. The mechanical mode shapes are identical except rotated by 90\si{\degree}, so their effective mass $m_\mathrm{eff}$ is the same (barring fabrication imperfections) and their response to thermal or white-noise driving should result in equal displacement (power) in both modes~\cite{Hauer2013}. Modelling this requires the linear decay rates, which we extract from the piecewise linear parts of the ringdown measurement, Fig.~1\textbf{b}. We obtain decay rates of $\gamma_1 = 2\pi\times 6.0$~\si{\hertz} and $\gamma_2 = 2\pi\times 0.7$~\si{\hertz} respectively for the peaks observed at $\omega_1 = 2\pi\times 240331.9$~\si{\hertz} and $\omega_2 = 2\pi\times 240341.1$~\si{\hertz}. 

Reconstructing the expected power spectrum results in the black lines of Fig.~\ref{Figspectrum}\textbf{b}. These suggest that instead of the detection efficiency (DE) being equal (dotted line), the lower-frequency peak is detected a factor 20 less efficiently (solid line). This is likely determined by the precise position of the laser beam on the membrane. From the (room-temperature) thermally driven amplitude ($\lesssim$\SI{2}{\pico\meter}), we deduce that the maximum driven amplitude $\simeq$\SI{2}{\nano\meter}, well within the linear regime of the interferometer. In that same comparison, no frequency shift of the peak is visible which confirms that we are in the linear regime of mechanical motion and the Duffing nonlinearity is small. The rounded shape of the peaks is determined by the minimum detection bandwidth of the spectrum analyzer, \SI{1}{\hertz}.

\subsection{Linear and nonlinear coupling}\label{Nonlinearcoupling}
\begin{figure*}
\includegraphics[width = \textwidth]{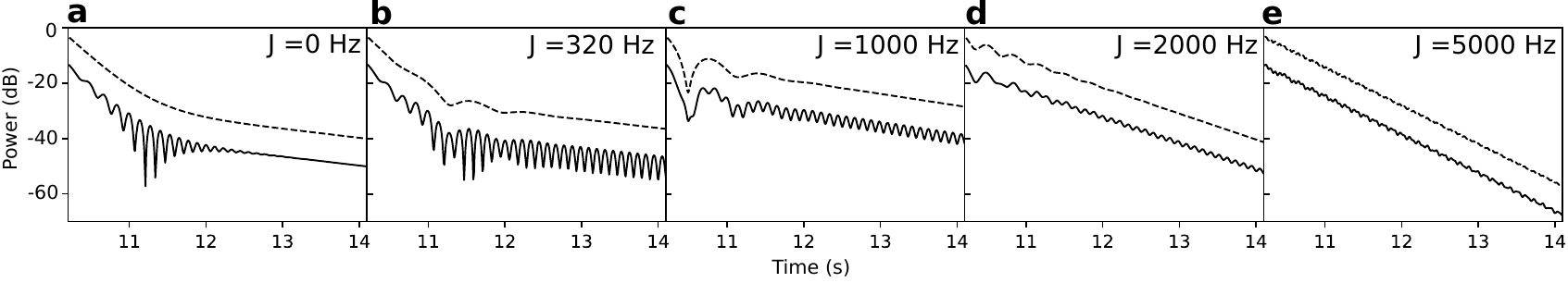}
\caption{Simulated ringdowns with (solid) and without (dashed) the beating pattern due to downmixing, for different values of coupling strength $J$. All simulations share the same initial values as Fig.~1\textbf{e} in the main text, the dashed lines are vertically offset by \SI{10}{\decibel} for clarity.}
\label{Figlinearcoupling}
\end{figure*}

\begin{figure*}
\includegraphics[width = \textwidth]{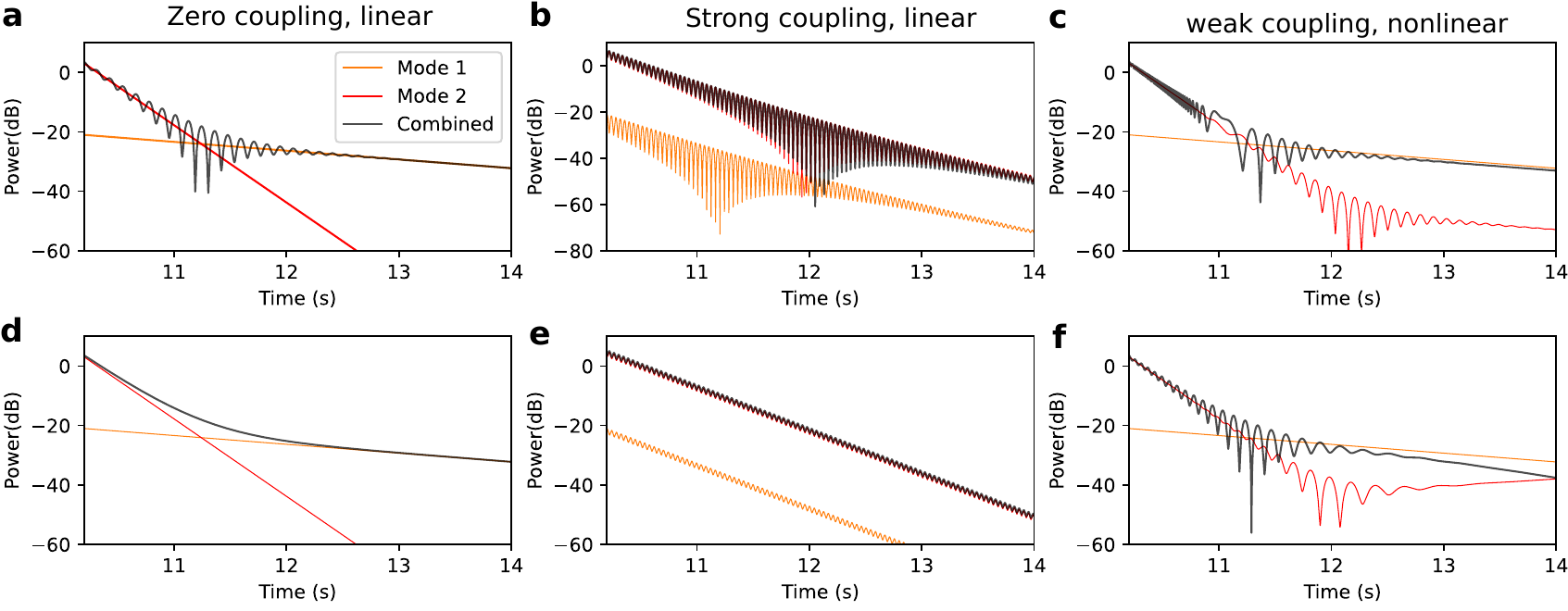}
\caption{Simulated ringdowns using the nonlinear coupled model of Ref.~\cite{Guttinger2017}. Top row (\textbf{a,b,c}) include the beating, equivalent to the detuning in the model of Ref.~\cite{Guttinger2017}, while the bottom row (\textbf{d,e,f}) excludes this effect. \textbf{a,d}: Ringdowns for zero coupling, replicating the result in Fig.~\ref{Figlinearcoupling}. \textbf{b,e}: Ringdowns in the strong, linear coupling regime for different values of coupling rate $J$. \textbf{e} matches to Fig.~\ref{Figlinearcoupling}, but \textbf{b} has an enhanced beat pattern due to the frequency splitting of the strong coupling. \textbf{c,f}: Ringdowns for nonlinear resonators, showing a beating pattern that slows down as the amplitudes decay. All simulations share the same initial values as Fig.~1\textbf{e} in the main text.}
\label{Fignonlinearcoupling}
\end{figure*}
To distinguish between the coupling and beating effects, we simulate and show ringdowns in Fig.~\ref{Figlinearcoupling} for various values of the coupling rate $J$. All simulations share the same initial condition as Fig.~1\textbf{e} in the main text. In Fig.~\ref{Figlinearcoupling}\textbf{a}, there is no coupling, $J = 0$. In the absence of the beating effect (dashed line), there is a smooth transition between two slopes. With the presence of beating effect (solid line), the kink between the slopes gains the characteristic pattern where the beat first grows in amplitude until the detected powers of the two peaks are equal, and then decreases in amplitude until it disappears. 

In Fig.~\ref{Figlinearcoupling}\textbf{b}, we replicate the ringdown of Fig.~1\textbf{e}. Without the beating effect (dashed line), there is some oscillation present that is indicative of energy exchange between the peaks. It appear superimposed on the beating pattern (solid line), illustrating that the effects are independent and both are necessary to describe the observations. 

In Figs.~\ref{Figlinearcoupling}\textbf{c-e}, the coupling rate is increased until the strong-coupling regime is reached ($J = 5000$~\si{\hertz} corresponds to a splitting of \SI{104}{\hertz} if $\omega_1 = \omega_2$). The oscillations due to the energy exchanged via the coupling increase in frequency, while the beating effect decreases in amplitude. The two different slopes also transform into one slope in the strong coupling regime. Here, energy predominantly leaves the system via the fast-decaying resonator. This illustrates that for strong enough coupling, the two slopes merge into one single slope.

The beating observed in ringdowns could also be related to energy exchange between nonlinear resonators with a 1:1 resonance~\cite{Guttinger2017}. The equations describing this are~\cite{Guttinger2017}
\begin{equation}
\begin{aligned}
\dot{a}_1 &= -i \left( \delta_1 a_1 - \frac{3}{2} \alpha_1 a_1 |a_1|^2 + J a_2\right) - \frac{\gamma_1}{2}a_1 \\ 
\dot{a}_2 &= -i \left( \delta_2 a_2 - \frac{3}{2} \alpha_2 a_2 |a_2|^2 + J a_1 \right) - \frac{\gamma_2}{2}a_2. 
\label{eqnlresonance}
\end{aligned}
\end{equation}
Here, $a_1, a_2$ are the complex mode amplitudes, $\delta_1, \delta_2$ describes the detuning of $a_1, a_2$ with respect to the rotating frame we choose (we pick $\delta_1 = 0, \delta_2 = 9$~\si{\hertz} to match the \SI{9}{\hertz} difference between our two peaks), $\alpha_1, \alpha_2$ are the Duffing nonlinearities, and $J$ and $\gamma_1, \gamma_2$ are the coupling and linear decay rates as in the main text.

We simulate our system using Eq.~\eqref{eqnlresonance}, and show the results in Fig.~\ref{Fignonlinearcoupling}. In the absence of coupling or nonlinearities ($J = 0, \alpha_1 = \alpha_2 = 0$), Eq.~\eqref{eqnlresonance} gives identical results to the model introduced in the main text. The detuning $\delta$ takes the role of the beating effect, if we plot $|a_1 + a_2|^2$. This illustrates the generality of the effect, as $a_1, a_2$ exist in a rotating frame (similar to the center frequency of our detector bandwidth). Figs.~\ref{Fignonlinearcoupling}\textbf{a,d} thus match closely with simulated results of Fig.~\ref{Figlinearcoupling}\textbf{a}. In the strong coupling regime (Figs.~\ref{Fignonlinearcoupling}\textbf{b,e}), there is a difference due to the large frequency split. The model introduced in the main text explicitly contains the frequencies observed from a measurement of the spectrum ($\omega_1, \omega_2$), and thus constrains the frequency difference to this value. In contrast, Eq.~\eqref{eqnlresonance} does not constrain this frequency splitting. Nonetheless, both models predict a single slope with fast, small oscillations. 

In Figs.~\ref{Fignonlinearcoupling}\textbf{c,f}, we introduce nonlinearity in the resonators ($\alpha_1 \neq 0, \alpha_2 \neq 0$). This creates an oscillating pattern, where the frequency of oscillation decreases as the resonators decay. This trend is general (i.e.~not dependent on the sign or values of $\alpha$), and different from the trends observed in the main text. In Fig.~1\textbf{e}, the beating pattern stays constant in frequency, while in Fig.~2\textbf{a}, it increases in frequency. Thus these observations are inconsistent with a model whereby the beating originates from the nonlinearity of the resonators. 

\begin{figure}
\includegraphics[width = 0.5\textwidth]{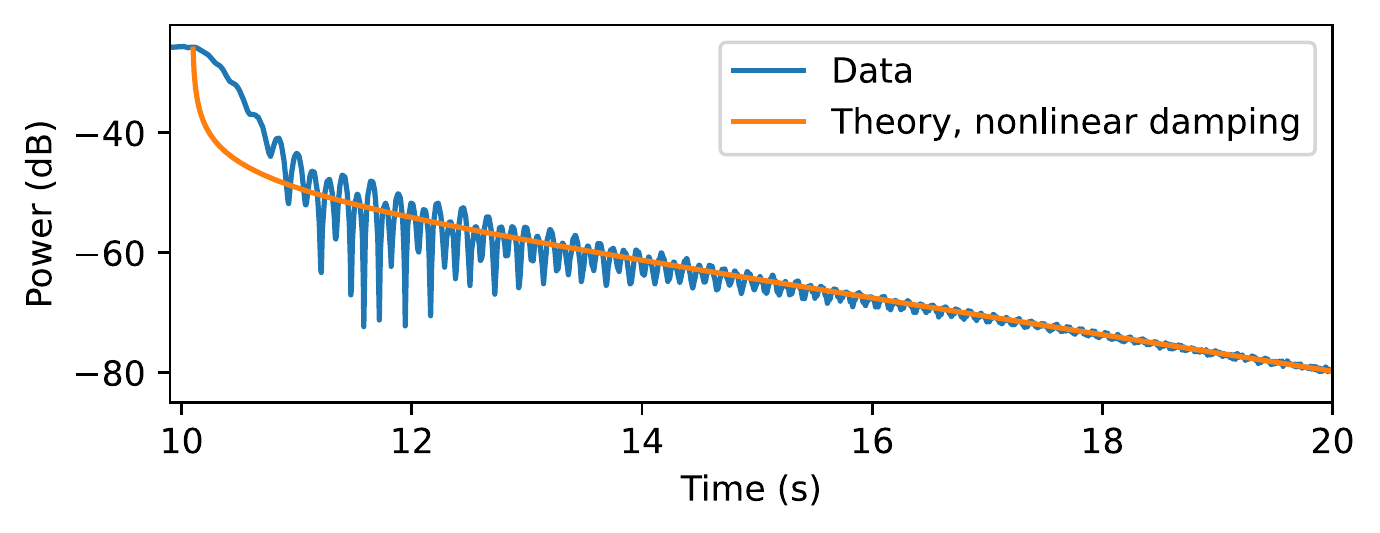}
\caption{Simulated ringdown (orange) for a resonator with nonlinear damping, with the observed ringdown (blue) for comparison.}
\label{Fignonlineardamping}
\end{figure}

Finally, the two linear slopes observed in the ringdowns could alternatively be explained by a single resonator that has nonlinear damping~\cite{Polunin2016}, of the form
\begin{equation}
\ddot{x} + 2(\gamma_\mathrm{l} + \gamma_\mathrm{nl} x^2) \dot{x} + \omega^2 x = 0.
\end{equation}
Here, $\gamma_\mathrm{l}$ the linear damping used in the main text and $\gamma_\mathrm{nl}$ the nonlinear damping. This term causes a fast decay directly after the driving stops, and returns the slow decay for lower amplitude. Qualitatively, the kink that results from this transition could have the same shape as the observed kink. However, the fast decay of the nonlinear damping region is much steeper than the fast decay we observe as show in Fig.~\ref{Fignonlineardamping}, which rules out this nonlinear damping as an alternative explanation.

\end{document}